# Ultrafast Spectroscopy of Bi$_2$Se$_3$ Topological Insulator

P. Sharma[1, 2,a)], D. Sharma[1, 2], N. Vashistha[1, 2], P. Rani[1], M. Kumar[1, 2], S. S. Islam[3], V.P.S. Awana[1, 2]

[1]*CSIR-National Physical Laboratory, K.S. Krishnan Marg, New Delhi-110012, India*
[2]*Academy council of scientific and innovative research, Ghaziabad U.P.-201002, India*
[3]*Centre of nanoscience and nanotechnology, Jamia Millia Islamia, New Delhi-110025, India*
[a)]: Corresponding Author- vatsprince8121994@gmail.com

**Abstract.** We investigate the ultrafast transient absorption spectrum of Bi$_2$Se$_3$ topological insulator. Bi$_2$Se$_3$ single crystal is grown through conventional solid-state reaction routevia self-flux method. The structural properties have been studied in terms of high-resolution Powder X-ray Diffraction (PXRD). Detailed Rietveld analysis of PXRD of the crystal showed that sample is crystallized in the rhombohedral crystal structure with a space group of R-3m, and the lattice parameters are a=b=4.14(2)Å and c=28.7010(7)A°. Scanning Electron Microscopy (SEM) result shows perfectly crystalline structure with layered type morphology which evidenced from surface XRD. Energy Dispersive Spectroscopy (EDS) analysis determined quantitative amounts of the constituent atoms, found to be very close to their stoichiometric ratio. Further the fluence dependent nonlinear behaviour is studied by means of ultrafast transient absorption spectroscopy. The ultrafast spectroscopy also predicts the capability of this single crystal to generate Terahertz (THz) radiations (T-rays).

## INTRODUCTION

Topological Insulator (TI) is the new state of matter in condensed matter physics which have a bulk bandgap and conducting surface states [1, 2]. Topological insulator has the properties of time-reversal symmetry (TRS) and spin-orbit coupling (SOC) [3]. A topological insulator is decided by Z2 invariants (v0; v1; v2; v3) which means special surface state exist in it [4]. Because of their unique properties, it may be useful for various applications in optoelectronics, spintronics,etc. [5]. The Bi$_2$Se$_3$ (Bismuth Selenide) is a strongTIhaving single Dirac point and a bandgap of 0.35eV [6]. Due to having the quintuple layer in the Bi$_2$Se$_3$ crystal makes it a good candidate for the thermoelectric material as well [7]. In addition, Bi$_2$Se$_3$ is a good candidate for the Terahertz (THz) applications including the high-frequency chip applications, tunable THz nonlinear optical devices [8]. In this report we investigate Bi$_2$Se$_3$ topological insulator for terahertz applications. This also computes the excited state dynamics in Bi$_2$Se$_3$ single crystal which has been further analyzed for the potential of this crystal to generate T-rays.

## EXPERIMENTAL

We are using conventional solid-state reaction route via self flux method to prepare Bi$_2$Se$_3$ single-crystal TI [9]. High purity (99.99%), Bi (Bismuth) and Selenium (Se) weretaken in their stoichiometric ratio for 1 gram sample. After weighing, the powders were ground properly in Ar filled Glove Box. Afterward, the mixture was pelletized with the help of hydraulic press (50kg/cm$^2$) followed by vacuum (10$^{-5}$ bar pressure) encapsulation in the quartz tube. Then the sealed sample was kept in tube furnace under a heat treatment is given in fig. 1(a). After a long heat treatment finally the sample is obtained in the form silvery and shiny crystal shown in fig. 1(b). The structural characterization of Bi$_2$Se$_3$single crystal was performed through room-temperature X-ray diffraction (XRD) using Rigaku made Mini Flex II of Cu-K$_α$radiation (λ=1.5418 Å). SEM study followed by EDS was performed on Bruker made scanning electron microscope. Raman study has been done to observe the vibrational modes of Bi$_2$Se$_3$ crystal using the Renishaw Raman Spectrometer.

However, the ultrafast transient absorption spectroscopy system used for measurement of decay profile of charge carriers uses a Ti: Sapphire based femtosecond laser oscillator (Micra by Coherent) and amplifier (Legend by Coherent). The output from the amplifier was a Gaussian pulse with a 60nm spectral width, 1kHz repetition rate, and 800nm center frequency. This beam is split into 2 parts through a 70:30 beam splitter. The higher power beam (pump) is used to excite the sample while the other portion of the beam (probe) is used to study the pump induce changes. Both the beams are made to fall on the sample at the same point inside the spectrometer (Helios). We can get a wide tunability (290nm-1600nm) in pump wavelength through the use of an optical parametric amplifier.

Through an optical chopper the repetition rate of pump is set at 500Hz for the measurement. The probe is stirred into the spectrometer through an 8ns long delay stage to ensure a time delay between arrival of pump and probe over the sample. Inside the spectrometer, probe is been converted into a white light continuum by the use of a sapphire/CaF$_2$ crystal. This enables a wide spectral range (320nm- 1600nm) for the study of optical behavior.

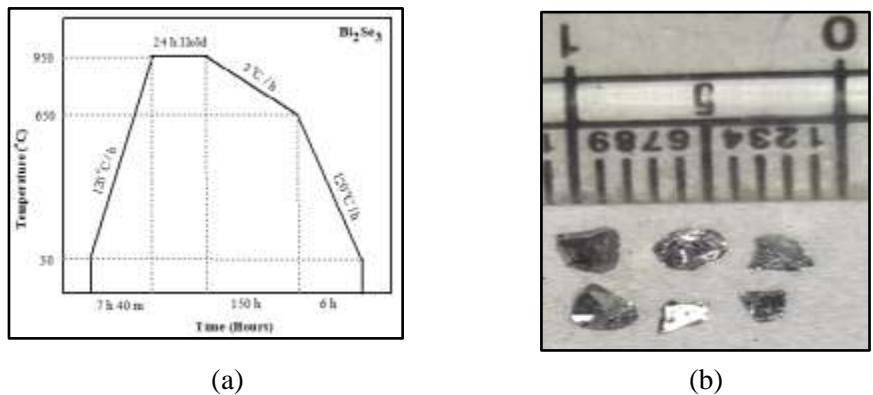

(a)                             (b)

**FIGURE 1** (a) Heat Treatment diagram of Bi$_2$Se$_3$ TI, and (b) Photograph of Bi$_2$Se$_3$ single crystal pieces

## RESULTS AND DISCUSSIONS

Figure 2 represents the Rietveld refinement of X-ray diffraction pattern of the powder form of Bi$_2$Se$_3$ crystal using the FullProf suite toolbar. From graph it is clear that the as grown Bi$_2$Se$_3$ crystal is crystallized in rhombohedral crystal structure with R-3m (D5) space group [9]. The lattice parameters and atom positions obtained from the Rietveld refinement area = 4.1793(5) A°, b = 4.1793(5) A° and c = 28.6183(2) A° and Bi (0, 0, 0.3999(5)), Se1 (0, 0, 0) and Se2 (0, 0, 0.2070(4)) respectively. Inset (a) of Fig. 2(A) shows the XRD pattern of the surface of flake of as grown crystal with growth of the crystal in [00*l*] plane only. Inset (b) of fig. 2(A) shows the Raman spectrum of bulk Bi$_2$Se$_3$ single crystal recorded at room temperature. Clearly, the spectrum shows three distinct Raman active modes at around 72.1, 131.2 and 177.1cm$^{-1}$ correspond to $A_{1g}^1$, $E_g^2$ and $A_{1g}^2$ respectively, which are in good agreement to the earlier reported results [10]. Figure 2(B) shows the unit cell structure of studied Bi$_2$Se$_3$ single crystal created using the VESTA software. The unit cell contains three bi-layers of Bi and Se stacked monolayers of either Bi or Se in a close-packed FCC structure.

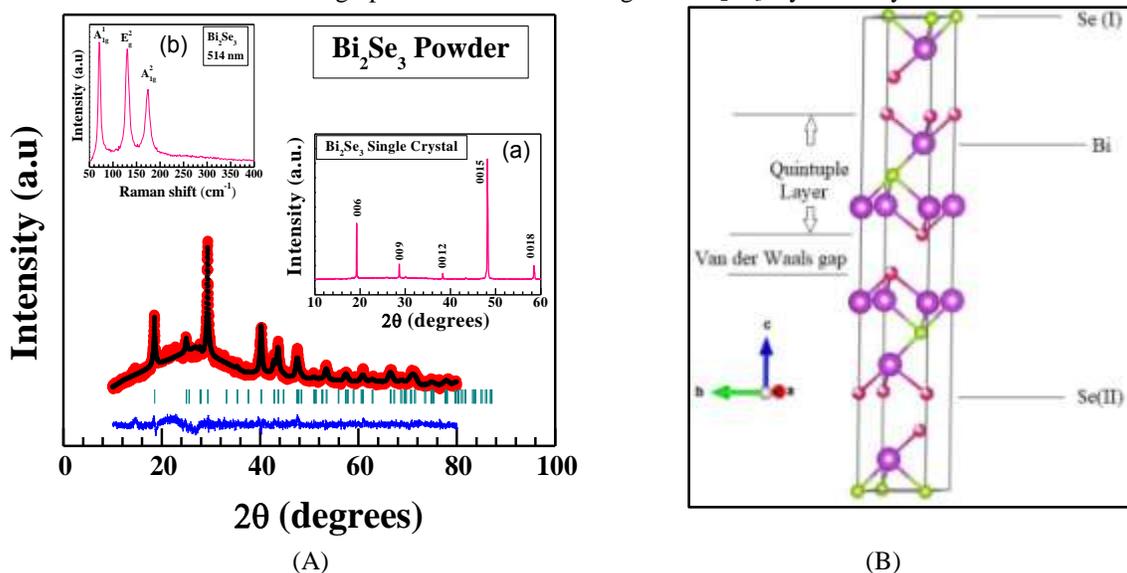

(A)                             (B)

**FIGURE 2.** (A) Rietveld fitted PXRD pattern of Bi$_2$Se$_3$ Singlecrystal, inset (a) is surface XRD pattern of the crystal flake and inset (b) is the Raman spectrum of Bi$_2$Se$_3$ crystal, (B) The unit cell of Bi$_2$Se$_3$ crystal, exhibiting Van der Waals gap in Bismuth(Bi) and Selenium (Se$_1$, Se$_2$).

Figure 3 (a,b) spectacles the SEM and EDS analysis of $Bi_2Se_3$ single crystal. These results show perfectly crystalline structure having layered type morphology which evidenced from single crystal XRD pattern. EDS analysis determined quantitative amounts of the constituent atoms, found to be very close to stoichiometric.

We have obtained the ultrafast transient absorption spectroscopy data using 650 nm as pump pulse and probed in the NIR region. The time-domain data have been collected up to 8ns. The data has been obtained at varying flux (0.5mW, 1.0 mW, 1.5 mW, and 2.0 mW average powers) to observe the non-linear behavior of the crystal. The spot size of the pump beam is about 400 micron and the repetition rate is 500Hz.

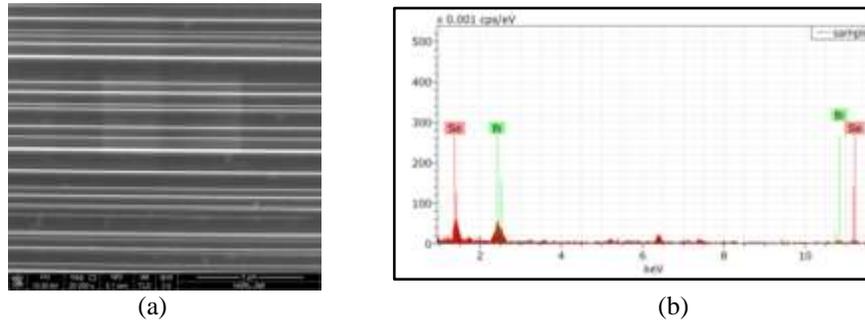

(a) (b)

**FIGURE 3.** (a) SEM image and (b) EDS of $Bi_2Se_3$ crystal showing layered structure and quantitative amounts of the constituent atoms.

The kinetic profiles contain the large oscillations and the large oscillations are further composed of the small oscillations which are sustained upto 10 ps. The small oscillations are due to the acoustic phonon components. These small oscillations are then converted to frequency domain. The kinetic profiles of the small oscillations obtained in time domain have been converted to frequency domain using fast Fourier transform (FFT). The FFT data gives the frequency domain spectrum. The frequency-domain data obtained using FFT are shown in Fig. 4.

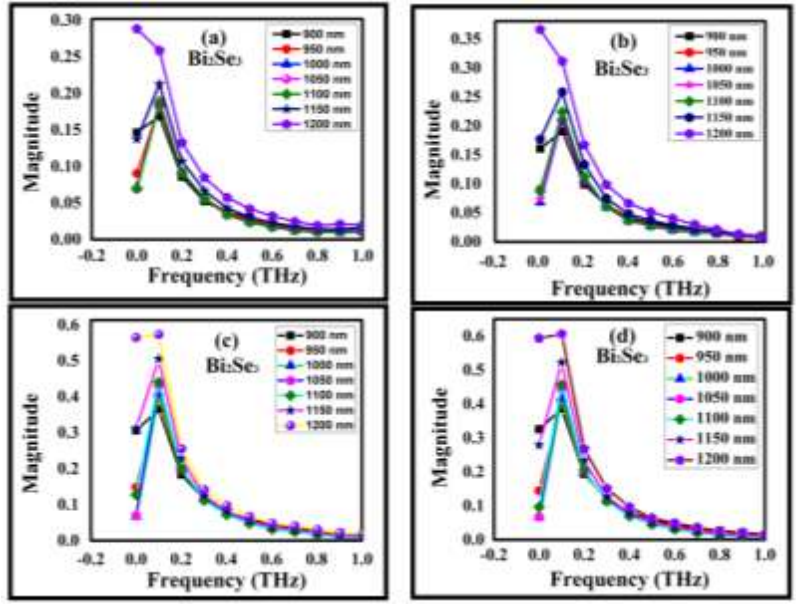

**FIGURE 4.** Wavelength dependent plot of the magnitude of THz with their generated terahertz frequency at an average power (a) 0.5 mW (b) 1.0 mW (c) 1.5 mW (d) 2.0 mW for $Bi_2Se_3$ crystal

The FFT data displays magnitude of the frequency components. Fig. 4 (a) shows the FFT of 650 nm pumped data up to 10ps with a fluence of 0.5mW. Fig. 4(b) shows the FFT profile of 1.0 mW, Fig. 4(c) shows the FFT data of 1.5 mW and Fig. 4(d) shows the FFT data of 2.0mW average powers. The sustained oscillations are of frequency ~0.1THz while the band of frequency varies with fluence slightly.

Figure 4 shows the magnitude of the THz frequencies variation with respect to the spectral variation and

Figure 5 represents the dependence of the magnitude with pump average power. Comparing Fig. 4 and Fig. 5 shows that while the sustained oscillations are up to 0.1 THz, the band keeps on varying with the fluence (average power). With all the fluences the frequency response band is up to 1.0 THz. While the magnitude of the sustained oscillations keeps on increasing with the fluence. This shows that sustained oscillations are more prominent at higher fluences. The fluence in this current study is limited to 2.0mW average power. However, it is predicted that with higher fluence the band and the magnitude may increase. This study is under progress and will be provided in a detailed manuscript.

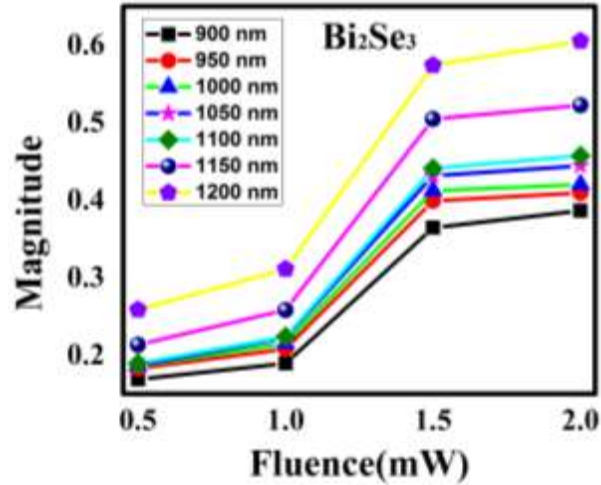

**FIGURE 5.** The dependence of the magnitude of THz with the pump average power.

## CONCLUSION

We have presented the ultrafast transient absorption spectroscopy of $Bi_2Se_3$ single crystal and have studied the non-linear response of the crystal in the various spectral ranges. While performing the frequency domain analysis of the time domain data shows that the crystal is capable of generating THz waves up-to 1THz (peak to 1/e value) at a fluence of 2.0 mW (average power). Further studies are in progress and will be presented in the detailed work.

## ACKNOWLEDGMENTS

The authors from CSIR-NPL would like to thank their Director NPL, India, for his keen interest in the present work. Authors further thank Dr. Bhasker Gahtori for XRD and Dr. S.S Islam for SEM. Prince Sharma thanks UGC, India, for research fellowship and AcSIR-NPL for Ph.D registration. Prince Sharma thanks K.M Kandpal for sealing samples in vacuum.